\documentstyle[12pt]{article}

\renewcommand{\theequation}{\thesection.\arabic{equation}}

\newlength{\extraspace}
\newlength{\extraspaces}
\newcounter{dummy}

\newcommand{\baa}{
\addtocounter{equation}{1}
\setcounter{dummy}{\value{equation}}
\setcounter{equation}{0}
\renewcommand{\theequation}{\thesection.\arabic{dummy}\alph{equation}}
\begin{eqnarray}
\addtolength{\abovedisplayskip}{\extraspaces}
\addtolength{\belowdisplayskip}{\extraspaces}
\addtolength{\abovedisplayshortskip}{\extraspace}
\addtolength{\belowdisplayshortskip}{\extraspace}}

\newcommand{\eaa}{
\end{eqnarray}
\setcounter{equation}{\value{dummy}}
\renewcommand{\theequation}{\thesection.\arabic{equation}}}

\newcommand{\be}{\begin{equation}
\addtolength{\abovedisplayskip}{\extraspaces}
\addtolength{\belowdisplayskip}{\extraspaces}
\addtolength{\abovedisplayshortskip}{\extraspace}
\addtolength{\belowdisplayshortskip}{\extraspace}}
\newcommand{\ee}{\end{equation}}

\newcommand{\ba}{\begin{eqnarray}
\addtolength{\abovedisplayskip}{\extraspaces}
\addtolength{\belowdisplayskip}{\extraspaces}
\addtolength{\abovedisplayshortskip}{\extraspace}
\addtolength{\belowdisplayshortskip}{\extraspace}}
\newcommand{\ea}{\end{eqnarray}}

\newcommand{\bd}{\begin{displaymath}
\addtolength{\abovedisplayskip}{\extraspaces}
\addtolength{\belowdisplayskip}{\extraspaces}
\addtolength{\abovedisplayshortskip}{\extraspace}
\addtolength{\belowdisplayshortskip}{\extraspace}}
\newcommand{\ed}{\end{displaymath}}

\newcommand{\ban}{\begin{eqnarray*}
\addtolength{\abovedisplayskip}{\extraspaces}
\addtolength{\belowdisplayskip}{\extraspaces}
\addtolength{\abovedisplayshortskip}{\extraspace}
\addtolength{\belowdisplayshortskip}{\extraspace}}
\newcommand{\ean}{\end{eqnarray*}}

\newcommand{\newsection}[1]{
\vspace{15mm}
\pagebreak[3]
\addtocounter{section}{1}
\setcounter{equation}{0}
\setcounter{subsection}{0}
\setcounter{footnote}{0}
\begin{center}
{\Large \thesection. #1}
\end{center}
\nopagebreak
\medskip
\nopagebreak}

\newcommand{\startappendix}{
\renewcommand{\thesection}{\Alph{section}}
\setcounter{section}{0}}

\newcommand{\newsubsection}[1]{
\vspace{1cm}
\pagebreak[3]

\addtocounter{subsection}{1}
\noindent{ \sc \thesubsection. #1}
\nopagebreak
\vspace{2mm}
\nopagebreak}

\newcommand{\nonu}{\nonumber \\[.5mm]}

\newcommand{\deel}[2]{{\textstyle{#1 \over #2}}}
\newcommand{\hf}{{\textstyle{1\over 2}}}

\newcommand{\ie}{{\it i.e.}}

\newcommand{\re}{\mbox{I}\!\mbox{R}}
\newcommand{\lha}{\left[}
\newcommand{\rha}{\right]}

\newcommand{\uit}{\wedge}
\newcommand{\lie}[1]{{\bf #1}}
\newcommand{\dif}{\partial}
\newcommand{\dbar}{\bar{\dif}}
\newcommand{\abar}{A_{\bar{z}}}
\renewcommand{\ll}{\Lambda}
\renewcommand{\ss}{\Sigma}
\newcommand{\tr}{\mbox{Tr}}
\newcommand{\del}{\delta}
\newcommand{\pri}{\Pi_i}
\newcommand{\prk}{\Pi_k}
\newcommand{\prid}{\Pi_i^{\dagger}}
\newcommand{\prkd}{\Pi_k^{\dagger}}
\newcommand{\pry}{\Pi_y}

\newcommand{\vp}{\varphi}
\newcommand{\ro}{\rho}
\newcommand{\rb}{\bar{\rho}}
\newcommand{\mub}{\bar{\mu}}
\newcommand{\nub}{\bar{\nu}}
\newcommand{\bg}{{\bf g}}
\newcommand{\act}[1]{\deel{k}{2\pi}\int d^2z \,\, \tr ( {#1} ) }
\newcommand{\wips}{W_{\psi}}
\newcommand{\abaps}{A^0_{\bar{z},\psi}}

\newcommand{\var}[1]{{\del \over \del {#1}}}
\newcommand{\vars}[2]{{\del {#1} \over \del {#2}}}

\newcommand{\adj}[1]{g^{-1} {#1} g}

\newcommand{\mat}[9]{\left( \begin{array}{ccc}
				#1 & #2 & #3 \\
				#4 & #5 & #6 \\
				#7 & #8 & #9
                             \end{array} \right) }

\newcommand{\cs}{Chern--Simons theory}

\newcommand{\sam}{Zamolodchikov}

\newcommand{\np}[1]{Nucl. Phys. {\bf B#1}}
\newcommand{\cmp}[1]{Comm. Math. Phys. {\bf #1}}
\newcommand{\intmod}[1]{Int. Journal of Mod. Phys. {\bf A#1}}
\newcommand{\plb}[1]{Phys. Lett. {\bf B#1}}

\begin{document}
\addtolength{\baselineskip}{.7mm}

\thispagestyle{empty}
\begin{flushright}
{\sc THU}-91/24\\
12/91
\end{flushright}
\vspace{1.5cm}

\begin{center}
{\LARGE\sc{$W$ Gravity From Chern--Simons Theory}}\\[2.5cm]

\sc{Jan de Boer\footnote{e-mail: deboer@ruunts.fys.ruu.nl}
and Jacob Goeree\footnote{e-mail: goeree@ruunts.fys.ruu.nl}}
\\[8mm]
{\it Institute for Theoretical Physics\\[2mm]
University of Utrecht\\[2mm]
Princetonplein 5\\[2mm]
P.O. Box 80.006\\[2mm]
3508 TA Utrecht}\\[3cm]

{\sc Abstract}\\[1.5cm]
\end{center}

\noindent
Starting with three dimensional \cs\ with gauge group
$Sl(N,\re)$, we derive an action $S_{cov}$ invariant
under both left and right $W_N$ transformations. We give
an interpretation of $S_{cov}$ in terms of anomalies, and
discuss its relation with Toda theory.

\vfill

\newpage

\newsection{Introduction}

A few years ago \sam\ made a systematic study of the possible
extensions of the Virasoro algebra \cite{sammy}. Besides the
extensions involving Kac-Moody and superconformal currents, he
found a new non-linear extension of the Virasoro algebra, based
on the occurrence of a spin-three field $W$. This field,
together with the usual stress energy tensor $T$, forms
the so-called
$W_3$ algebra, which is defined by the following operator
product expansions:
\ba \label{w3alg}
T(z)T(w) &\sim& \frac{c/2}{(z-w)^4}+\frac{2T(w)}{(z-w)^2}+
\frac{\dif T(w)}{z-w},\nonu
T(z)W(w) &\sim& \frac{3W(w)}{(z-w)^2}+\frac{\dif W(w)}{z-w}, \nonu
W(z)W(w) &\sim& \frac{c/3}{(z-w)^6}+\frac{2T(w)}{(z-w)^4}+
\frac{\dif T(w)}{(z-w)^3}\nonu
& &+\frac{1}{(z-w)^2}\left( \deel{3}{10}\dif^2 T(w) +2b^2 \Lambda(w)
\right) \nonu
& &+\frac{1}{z-w} \left( \deel{1}{15} \dif^3 T(w)+b^2 \dif
\Lambda (w) \right),
\ea
where the non-linear term $\Lambda(z)$ is defined as
\be \label{lam}
\Lambda(z)=\, :T(z)T(z):-\deel{3}{10}\dif^2 T(z),
\ee
and the constant $b^2$, determined by associativity, reads
\be \label{b}
b^2=\frac{16}{22+5c}.
\ee

The above algebra can be generalized to the case of a $W_N$ algebra,
which contains fields $W^{(k)}$ of spin $k$, with $k$ running from 2
to $N$. In \cite{sander} it was shown that this algebra is
intimately related to the affine Kac-Moody algebra associated to
$su(N)$, in the sense that the fields $W^{(k)}$ can be constructed from
the Kac-Moody currents by using the higher-order Casimir invariants
of the underlying finite dimensional Lie algebra.
Besides this connection
with affine Kac-Moody algebras, it is now clear that these non-linear
extensions of the Virasoro algebra play a central role in many
other areas of two dimensional physics. They have been shown to appear
in Toda theories \cite{toda}, gauged WZW models \cite{wzw},
reductions of the KP hierarchy \cite{ds,classicalw}
and in the matrix model formulation of two dimensional
quantum gravity \cite{ijdel}.

Despite the relevance of these $W$ algebras for the above mentioned
branches of two dimensional physics, which makes it clear that they
represent some universal structure, it is fair to say that some
aspects of these algebras are still poorly understood.
Although by now we have many different realizations of these algebras
and most of the algebraic aspects are well sorted out, it is the
geometrical interpretation of these algebras which is still missing.
Whereas we know that the Virasoro algebra arises after gauge fixing
the two dimensional diffeomorphism invariance (conformal gauge),
a similar geometrical interpretation for the $W$ case is lacking.

In this paper we will try to uncover
some of the mysteries of `$W$ geometry'
by constructing an action $S_{cov}$ which has
{\em local} $W$ transformations
as its symmetries. This action then describes what is commonly denoted
as $W$ gravity. $S_{cov}$ can be viewed
as the $W$ generalization of the
covariant action for pure gravity,
first constructed by Polyakov \cite{pol}:
\be \label{poly}
S=\frac{c}{96\pi} \int \!\!\! \int R\, \Box^{-1} R.
\ee
We will construct $S_{cov}$ by starting with a topological gauge
theory in three dimensions, namely \cs\ \cite{ed}, on a three manifold
of the form $M = \Sigma \times \re$, where
$\Sigma$ is a two dimensional Riemann surface.
What we will end up with is then a theory for $W$ gravity
on $\Sigma$. The reason for this approach is that we believe that the
moduli space of $W_N$ gravity is related (if not equal) to the moduli
space of flat $Sl(N,\re)$ bundles. This latter space appears naturally
as the classical phase space of $Sl(N,\re)$ \cs, which hints towards a
possible connection between this theory and $W$ gravity. Indeed, it was
shown by H. Verlinde that the above action for
pure gravity (\ref{poly})
can be obtained from $Sl(2,\re)$ \cs\ \cite{herman}.

There have been previous constructions of actions which admit local
$W_3$ symmetries, which all amount to gauging of the
$W_3$ transformations. This technique has led to both
a chiral action and to a fully covariant action for $W_3$ gravity
\cite{kj2}. As we will show, our construction
results in an action closely related to the one of \cite{kj2}.
An important advantage of our
method is that we can apply it quite generally to $Sl(N,\re)$,
resulting in the covariant action for $W_N$ gravity. Specializing
to the case of $Sl(3,\re)$ gives our result for $W_3$ gravity,
which was already reported in \cite{jj}. Another advantage is
that formulating the $W_N$ gravity in terms of $Sl(N,\re)$ \cs\
should make it more tractable to study its moduli space.

This paper is organized as follows. In section 2 we will
review some generalities about \cs. We will introduce the
concepts needed for the construction of the covariant action,
such as wavefunctions, polarization, the inner-product, the
K\"{a}hler potential etc. Furthermore, we will illustrate our
method by first reviewing the construction of the covariant
action in the so-called `standard polarization'
\cite{beroog,witten2}. Next, in sections 3 and 4,
we will compute the covariant
action $S_{cov}$ for a different choice of polarization, namely
the choice which relates $Sl(N,\re)$ transformations to $W_N$
transformations. It will turn out that this
first result for the covariant action admits a large symmetry
group, which can be used to gauge away some of the degrees of
freedom. In addition $S_{cov}$ contains a number of auxiliary
fields which can be eliminated by replacing them by their
equations of motion. Details will be given for the case of
ordinary and $W_3$ gravity. In section 5 we will prove the
invariance of $S_{cov}$ under left and right $W_N$
transformations, discuss its relation with Toda theory,
and give an interpretation of $S_{cov}$ in terms of anomalies.
Finally, we will address some open problems and give some
concluding remarks in section 6.

\newsection{\cs}

\cs\ on a three manifold $M$ is described by the action
\be \label{cs}
S = \frac{k}{4\pi i} \int_{M} \tr(\tilde{A}\uit \tilde{d}\tilde{A}
+\deel{2}{3}\tilde{A}\uit \tilde{A}\uit \tilde{A}),
\ee
where the connection $\tilde{A}$ is a one form with values in the Lie
algebra \lie{g} of some Lie group $G$, and $\tilde{d}$ denotes the
exterior derivative on $M$. In this paper $M$ will be
of the form $M=\ss \times \re$, $\ss$ being a
Riemann surface, for which
$\tilde{A}$ and $\tilde{d}$ can be
decomposed into space and time components,
\ie\ $\tilde{A}=A_0dt+A$, with $A=A_zdz+A_{\bar{z}}d\bar{z}$,
and $\tilde{d}=dt\dif / \dif t+d$.
Rewriting the action as
\be \label{cs2}
S = \frac{k}{4\pi i} \int dt \int_{\ss}
\tr(A\uit \dif_t A + 2A_0(dA+A\uit A)),
\ee
we recognize that $A_0$ acts as a Lagrange multiplier which
implements the constraint $F=dA +A\uit A=0$. Furthermore, we
deduce from this
action the following non-vanishing Poisson brackets
\be \label{poisbracket}
\{ A^a_{\bar{z}}(z),A^b_{z}(w) \}=
\frac{2 \pi i}{k}\eta^{ab}\del(z-w),
\ee
where $A_{z}=\sum_a A^a_{z}T^a$, with
$\tr(T^aT^b)=\eta^{ab}$.

Upon quantizing the theory we have to replace the above
Poisson bracket by a commutator, and we have to choose a
`polarization.' This simply means that we have to divide
the set of variables
$(A^a_{z},A^a_{\bar{z}})$ into two subsets.
One subset will contain fields $X_i$ and
the other subset will consist of derivatives
$\var{X_i}$, in accordance with (\ref{poisbracket}).
The choice of these subsets is called
a choice of polarization.
We will denote the subset containing the fields by $\pri A_z$
and $\prk \abar$, where $\pri$ and $\prk$ are certain
projections on subspaces of the Lie algebra $\bg$. The standard
polarization corresponds to the case where $\pri={\bf 1}$ and
$\prk=0$.

Of course we also have to incorporate
the Gauss law constraints $F(A_z,A_{\bar{z}})=0$. Following
\cite{herman,moore,witten1} we will impose these constraints
{\em after} quantization. So we will first consider a `big'
Hilbert space obtained by quantization of
(\ref{poisbracket}), and then select the physical
wavefunctions $\Psi$
by requiring $F(A_z,A_{\bar{z}})\Psi=0$.

As an example of this construction let us consider the
standard polarization, in which the set of fields is
given by the $A^a_z$, so the $A^a_{\bar{z}}$ act as derivatives
$-\deel{2\pi}{k}\var{A^a_z}$. This implies that
the physical wavefunctions will be functions of $A_z$.
For this choice of polarization
it is well known \cite{beroog} that the solution of the
zero-curvature constraint
\be
F\left( A_z,\deel{2\pi}{k}\var{A_z}\right) \Psi(A_z)
=-:\deel{2\pi}{k}\dif \var{A_z}
+\dbar A_z + \deel{2\pi}{k}\, [ A_z, \var{A_z} ] : \Psi(A_z)=0,
\ee
is given by:
\be \label{psi}
\Psi(A_z)=\exp S(A_z) = \exp -S_{WZW}(g),
\ee
where $S_{WZW}(g)$ is the Wess-Zumino-Witten action:
\be \label{wzw}
S_{WZW}(g)=\deel{k}{4\pi}\int d^2z\,\, \tr(g^{-1}\dif g g^{-1}
\dbar g) - \deel{k}{12\pi} \int_B \tr(g^{-1}dg)^3,
\ee
and $A_z$ and $g$ are related via $A_z=g^{-1}\dif g$.

If we now want to compute transition amplitudes between
some initial physical state $\Psi_1$ and some final state
$\Psi_2$, we have to consider the inner-product between
these states in \cs. (The evolution operator is simply the
identity here, as the Hamiltonian vanishes for a topological
theory.) The expression for such an inner-product is:
\be \label{inprod}
\left< \Psi_1 \mid \Psi_2 \right> = \int DA e^{K(A_z,A_{\bar{z}})}
\bar{\Psi}_1(A_{\bar{z}}) \Psi_2(A_z).
\ee
This formula should be read as follows. (i) $DA$ is short for
$DA_zDA_{\bar{z}}$. (ii) $\bar{\Psi}(A_{\bar{z}})$ is the solution
of the zero-curvature constraint, but now with the role of
$A_z$ and $A_{\bar{z}}$ interchanged. So it is given by:
\be
\bar{\Psi}(A_{\bar{z}})=\exp \bar{S}(A_{\bar{z}}) = \exp -S_{WZW}(h),
\ee
with $A_{\bar{z}}=h \dbar h^{-1}$. (iii) The K\"{a}hler term
$K(A_z,A_{\bar{z}})$ appears since we want to take the inner-product
between wavefunctions depending on different variables, $A_z$
and $A_{\bar{z}}$, which are conjugate variables. So we should
perform a `Fourier' transformation. In (\ref{inprod}) this is
automatically taken care off if one does the integral over
$A_{\bar{z}}$ (or $A_{z}$), provided that the K\"{a}hler term
takes the form:
\be \label{kahler}
K(A_z,A_{\bar{z}})=\deel{k}{2\pi}\int d^2z \,\, \tr(A_zA_{\bar{z}}).
\ee
The inner-product can now be written as:
\be \label{inprod2}
\left< \Psi_1 \mid \Psi_2 \right> = \int DA
\exp S_{cov}(A_z,A_{\bar{z}}),
\ee
where $S_{cov}(A_z,A_{\bar{z}})=S(A_z)+\bar{S}(A_{\bar{z}})+
K(A_z,A_{\bar{z}})$ is a {\em covariant action}, \ie\
invariant under both left and right transformations given by
\ba \label{transf}
\del A_z &=& \dif \eta + [A_z,\eta], \nonu
\del A_{\bar{z}} &=& \dbar \eta + [A_{\bar{z}},\eta].
\ea

Expressed in terms of the group variables $g,h$ $S_{cov}$
takes the following form:
\be \label{wzwcov}
S_{cov} = -S_{WZW}(g)-S_{WZW}(h)-\deel{k}{2\pi}\int d^2z\,\,
\tr(g^{-1}\dif g \dbar h h^{-1}).
\ee
This action is invariant under the transformations $g\rightarrow
gf,\,h\rightarrow f^{-1}h$, which can be easily proven if one
makes of the Polyakov--Wiegman formula \cite{powie}:
\be \label{powi}
S_{WZW}(gf)=S_{WZW}(g)+S_{WZW}(f)+\deel{k}{2\pi}\int d^2z\,\,
\tr(g^{-1}\dif g  \dbar f f^{-1}).
\ee
(In fact the invariance of $S_{cov}$ under $g\rightarrow gf,\,
h\rightarrow f^{-1}h$ is just the integrated form of (\ref{transf}).)
{}From this invariance one suspects that it should be possible
to write $S_{cov}$ in terms of the invariant product $G=gh$.
Indeed, comparing (\ref{wzwcov}) with (\ref{powi}), it
is evident that the covariant action is given by
\cite{pol2,witten2}:
\be \label{finalwzw}
S_{cov}=-S_{WZW}(G).
\ee

In this paper we will repeat the above steps for a different
polarization, namely one which relates the $Sl(N,\re)$ gauge
transformations to $W_N$ transformations. For this polarization we
will solve the Gauss law constraint $F\Psi=0$, compute the
inner-product between two wavefunctions
$\Psi_1=\exp S$ and $\bar{\Psi}_2=\exp \bar{S}$,
add a K\"{a}hler term $K$, to end up with a covariant action
$S_{cov}=S+\bar{S}+K$. $S_{cov}$ will depend on
two group variables $g,h$,
which are elements of a certain subgroup of $Sl(N,\re)$,
and on $2(N-1)$ parameters $\mu_i,\bar{\mu}_i$ ($i=2,\ldots ,N$)
which will play the role of conjugate variables of the $W_i,\bar{W}_i$
fields of the $W_N$ algebra.
We will show that for this non-standard polarization
$S_{cov}$ is again invariant under transformations
of the form $g\rightarrow gf,\, h\rightarrow f^{-1}h$,
where $f$ is now restricted to a $Sl(N-1,\re)\times \re$ subgroup of
$Sl(N,\re)$. Using this invariance we will be able to
rewrite the action in terms of the invariant product $G=gh$. The
resulting action $S_{cov}(G)$ is our definition for the covariant
action for $W_N$ gravity. This action is invariant under both left and
right $W_N$ transformations. So, in comparison to the standard
polarization, we have somehow split the $Sl(N,\re)$ symmetry
transformations into $Sl(N-1,\re)\times \re$ and
$W_N,\bar{W}_N$ transformations.
Note that this splitting is in agreement with the following
dimension formula
\be \label{opendeurbull}
\mbox{dim}\,sl(N,\re)=\mbox{dim}\,sl(N-1,\re)+1+2(N-1).
\ee

\newsection{The Solution of the Zero-Curvature Constraint}

In this section we will solve the zero-curvature constraints of
\cs\ for $G=Sl(N,\re)$
in a certain, nonstandard polarization, following closely
the strategy of \cite{jj}. Let us first state the main result of
this section: in the polarization where we take as fields $\pri
A_z$ and $\prk \abar$, and as derivatives with respect to these
fields therefore $\prkd A_z$ and $\prid \abar$, the solutions of
the zero-curvature constraints are wave functions of the form:
\be \label{solution}
\Psi[\pri A_z,\prk \abar]=e^{S(\pri A_z,\prk
\abar)} \Psi[\mu_2,\ldots,\mu_N],
\ee
where $\mu_k$ denote the $(1-k,1)$ differentials that occur
naturally in $W_N$-gravity (for instance, $\mu_2$ is just the
well-known Beltrami-differential), and
$\Psi[\mu_2,\ldots,\mu_N]$ solves the Ward identities of the
classical $W_N$-algebra \cite{classicalw}. The action $S$ is
given in equation (\ref{finals}). The nonstandard
polarization is given by the projections $\pri$ and $\prk$.
These are projections on certain subspaces of the Lie algebra
$sl(N,\re)$, that form closed sub-Lie algebras. The subalgebra
$\prk sl(N,\re)$ is the abelian subalgebra that consists of all
$N\times N$ matrices $M_{ij}$ with $M_{ij}=0$ unless $i<N$ and
$j=N$. The subalgebra $\pri sl(N,\re)$ consists of all traceless
$N \times N$ matrices $M_{ij}$ with $M_{ij}=0$ if $i=N$ and
$j<N$. More explicitly:
\be  \label{projdef}
\pri sl(N,\re)  =  \left( \begin{array}{cccc} * & \cdots & * &
* \\ \vdots &  & \vdots
& \vdots \\ {*} & \cdots & * & * \\ 0 & \cdots & 0 & *
\end{array} \right) ,\hspace{5mm}
\prk sl(N,\re)  = \left( \begin{array}{cccc} 0 & \cdots & 0 & * \\
\vdots & & \vdots & \vdots \\ 0 & \cdots & 0 & * \\ 0 & \cdots &
0 & 0 \end{array} \right).
\ee
The projections $\prid$ and $\prkd$ are defined through
$\prid={\bf 1}-\prk$ and $\prkd={\bf 1}-\pri$. Note that for arbitrary
$X,Y\in sl(N,\re)$ we have $\tr (\pri X \prk Y)=0$, so
that the Poisson bracket (\ref{poisbracket})
of any two fields is indeed zero. Another important property is
that
\be \label{prop1}
X\in \pri \bg,\, Y\in\prk\bg \Rightarrow [X,Y]\in\prk\bg.
\ee
In order to understand why we
need solutions of the classical $W_N$-Ward identities in
(\ref{solution}), we will first investigate the relation between
the zero curvature equation $F(A_z,\abar)=0$ and the classical
$W_N$-Ward identities.

\newsubsection{The $W_N$-Ward identities}

The relation between $W_N$-Ward identities and zero-curvature
equations is essentially due to Drinfel'd and Sokolov \cite{ds},
who showed that taking a particular form for the connection
$A_z$ gives in a natural way (via Hamiltonian reduction)
the second Gelfand-Dickii bracket \cite{gelfdick}. In turn, it
is known \cite{classicalw} that these brackets reproduce exactly
the classical form of the operator expansions of $W_N$-algebras.
It is therefore a natural starting point to take the same form
for $A_z$ as Drinfel'd and Sokolov did:
\be \label{azero}
A^0_z = \left( \begin{array}{cccccc} 0 & 1 & 0 & \cdots & 0 & 0
\\ 0 & 0 & 1 & \cdots & 0 & 0 \\ \vdots & \vdots & \vdots & & \vdots
& \vdots \\ 0 & 0 & 0 & \cdots & 1 & 0
\\ 0 & 0 & 0 & \cdots & 0 & 1 \\ W_N & W_{N-1} & W_{N-2} & \cdots &
W_2 & 0 \end{array} \right) .
\ee
We will write this also as
\be A^0_z = \Lambda + W \label{azerotwo} \ee
where $\ll$ denotes the matrix with only the one's next to the
diagonal, and $W$ denotes the piece containing only the fields
$W_i$, so that $W\in \prkd {\bf g}$. We will sometimes also
write $\ll=\sum_{i=1}^{N-1} e_{i,i+1}$, where $e_{i,j}$ is the
matrix with a one in its $(i,j)$ entry, and zeroes everywhere
else. The zero-curvature equation now reads
\be \label{zerocur}
F=\dif \abar^0 - \dbar W +[\ll+W,\abar^0]=0.
\ee
Since $A_z$ and $\abar$ are conjugate variables with respect
to the Poisson bracket (\ref{poisbracket}), and $A^0_z$ still
contains arbitrary fields $W_i$, we will put the fields conjugate
to the $W_i$ in $\abar$. These fields are precisely the
$(1-k,1)$ differentials $\mu_k$, and we therefore require that
\be \label{projabarzero}
\prk \abar^0 = \sum_{i=1}^{N-1} \mu_{N+1-i} e_{i,N}.
\ee
In matrix notation this means that $\abar^0$ has the form
\be \label{abarzero2}
A^0_{\bar{z}} =
\left( \begin{array}{cccc} {*} & \cdots & * & \mu_N \\
{*} & \cdots & * & \mu_{N-1} \\ \vdots & & \vdots & \vdots \\
{*} & \cdots & * & \mu_2 \\ {*} & \cdots & * & * \end{array}
\right) .
\ee
It is now possible to solve the equation $\pri F=0$ and to
determine all other entries in $\abar^0$. All other entries
in $\abar^0$ are polynomials in $\mu_i$, $W_i$, and their
derivatives. What remains are the $N-1$ equations $\prkd F=0$,
and these are precisely the Ward-identities of the
$W_N$-algebra, which follows from \cite{ds}. The explicit form
of $\abar^0$ is given in the appendix. For example, in the case
of $Sl(2,\re)$ one finds
\be
\abar^0=\left(\begin{array}{cc}
\deel{1}{2}\dif \mu & \mu \\
\mu T -\deel{1}{2} \dif^2 \mu & -\deel{1}{2} \dif \mu
\end{array}\right),
\ee
and the remaining zero-curvature equation reads
\be \label{ward1}
0 = F_{21}=-\deel{1}{2}\dif^3 \mu +\dif(\mu T)+(\dif\mu )T-\dbar T.
\ee
Here we used the standard notation $W_2=T$.

In \cs\ we have, however, an arbitrary connection $A$, and
not a special one like (\ref{azero}). It turns out that we need
gauge transformations to obtain $W_N$-Ward identities from
arbitrary connections $A$.

\newsubsection{The Role of Gauge Transformations}

If we impose zero-curvature constraints on wave-functions
$\Psi$, then we must specify (as in ordinary quantum mechanics)
how we quantize the expression
$F(A_z,\abar)=\dif\abar-\dbar A_z+[A_z,\abar]$, if we replace
$\prid \abar$ and $\prkd A_z$ by functional derivatives with
respect to $\pri A_z$ and $\prk \abar$. We will simply put all
derivatives to the right of the fields\footnote{A different
choice would differ from ours by terms that are of higher order
in $1/c$; because our approach is valid only up to the lowest
order in $1/c$, in which case the quantum $W_N$-algebra reduces
to the classical one, we can completely neglect such
differences.}. Looking at the expression for the Poisson bracket
(\ref{poisbracket}), we see that, when acting on wave
functions $\Psi[\mu_2,\ldots,\mu_N]$, $W_i$ should be identified
with $\frac{2\pi}{k}\var{\mu_i}$. The statement that
$\Psi[\mu_2,\ldots,\mu_N]$ solves the $W_N$-Ward identities, is
equivalent to
\be F(A_z^0,\abar^0)\Psi[\mu_2,\ldots,\mu_N]=0 \label{ward3}.
\ee
At this point we make the crucial observation, that if $g$ is an
arbitrary $Sl(N,\re)$-valued function, independent of the
$\mu_i$, equation (\ref{ward3}) implies
\be F((A_z^0)^g,(\abar^0)^g)\Psi=g^{-1}F(A_z^0,\abar^0)g\Psi=0
\label{gaugetrafo}, \ee
where $(A_z^0)^g$ and $(\abar^0)^g$ denote the gauge transformed
connections $g^{-1}A_z^0 g + g^{-1}\dif g$ and $g^{-1}\abar^0 g
+g^{-1}\dbar g$. We will assume that this is the most general
curvature one can write down, which annihilates those $\Psi$
that solve the $W_N$-Ward identities.

\newsubsection{The Solution}

Let us now go back to the original problem, that is, solving
the zero-curvature constraint $F\psi=0$, where
\be \prkd A_z= \frac{2\pi}{k}\var{\prk\abar} \ee and
\be \prid\abar= -\frac{2\pi}{k}\var{\pri A_z} , \ee
and let us look for solutions of type (\ref{solution}), \ie\
$\psi=e^S\Psi$. Multiplying the zero-curvature equation with
$e^{-S}$, it reads $(e^{-S}Fe^S)\Psi=0$. This equation can also
be written as
\be \label{zerocurv}
F\left( A_z+ \frac{2\pi}{k}\vars{S}{\prk\abar} ,\abar
-\frac{2\pi}{k}\vars{S}{\pri A_z}\right) \Psi=0. \ee
$F$ contains, in general, double derivatives, giving rise to
terms $\left(\frac{2\pi}{k}\right)^2 \frac{\delta^2 S}{\delta
A_1 \delta A_2}$ when working out $e^{-S}Fe^S$. However, these
terms are of higher order in $1/c$, and, as discussed
previously, we will ignore these. Now we want to show that
any $\Psi$ that solves the $W_N$-Ward identities, is a solution
of (\ref{zerocurv}). Because we assumed that the most general
curvature which annihilates such wave functions $\Psi$ is given
by (\ref{gaugetrafo}), we find, upon comparing
(\ref{gaugetrafo}) and (\ref{zerocurv}), that solutions of the
form (\ref{solution}) exist if and only if we can find an $S$
such that
\ba
A_z+ \frac{2\pi}{k}\vars{S}{\prk\abar} & = & (A^0_z)^g +
\mbox{{\rm derivatives not containing}}\,\,\, \var{\mu_i},
\label{eqn1} \\
\abar -\frac{2\pi}{k}\vars{S}{\pri A_z} & = & (\abar^0)^g+
\mbox{{\rm derivatives not containing}}\,\,\, \var{\mu_i}.
\label{eqn2}
\ea
Restricting (\ref{eqn1}) to the part in $\pri\bg$, we find that
$\pri A_z=\pri(g^{-1}\dif g+g^{-1}\ll g+g^{-1}Wg)$. The matrix
$W$ contains derivatives with respect to the $\mu_i$, and we
do not want derivatives in the parametrization of our fields.
Therefore, $g$ should satisfy $\pri(g^{-1}Wg)=0$. One may
easily verify that this restricts $g$ to be an element of
$G_P=\exp(\prid \bg)$, the group which has $\prid\bg$ as its
Lie algebra. This shows that we must parametrize $\pri A_z$ via
\be \label{par1}
\pri A_z = \pri(g^{-1}\dif g+g^{-1}\ll g)\equiv \pri \Lambda^g,
\ee
where we defined $\Lambda^g=g^{-1}\Lambda g +g^{-1}\dif g$.
Similarly, taking the $\prk$ of (\ref{eqn2}), we find the
parametrization for $\prk \abar$:
\be \label{par2} \prk\abar = \prk(g^{-1}\abar^0 g +
g^{-1}\dbar g ) = \prk(g^{-1}\abar^0 g).
\ee
Due to the fact that $\prk(g^{-1}\abar g)=\prk(g^{-1}(\prk
\abar^0)g)$, no derivatives will enter in the definition of
$\prk \abar$ either, and we conclude that we have a
parametrization of the $N^2-1$ fields $\pri A_z$ and $\prk
\abar$ in terms of $N^2-1$ independent variables, the $N^2-N$
components of $g\in G_P$, and the $N-1$ variables $\mu_i$. The
equations, which we still have to solve, are the components of
(\ref{eqn1}) in $\prkd\bg$, and the components of (\ref{eqn2})
in $\prid\bg$. They read, when separated into pieces which do
and do not contain derivatives,
\ba
\frac{2\pi}{k}\vars{S}{\prk\abar} & = & \prkd\Lambda^g, \label{sig1}\\
-\frac{2\pi}{k}\vars{S}{\pri A_z} & = &
\prid(g^{-1}A^0_{\bar{z},f}g+g^{-1}\dbar g ), \label{sig2} \\
\frac{2\pi}{k}\var{\prk\abar} & = & \prkd(g^{-1}Wg) +
\mbox{{\rm derivatives not containing}}\,\,\, \var{\mu_i},
\label{extr1} \\
-\frac{2\pi}{k}\var{\pri A_z} & = &
\prid(g^{-1}A^0_{\bar{z},d}g) +
\mbox{{\rm derivatives not containing}}\,\,\, \var{\mu_i},
\label{extr2}
\ea
where $\abar^0=A^0_{\bar{z},f}+A^0_{\bar{z},d}$, and
$A^0_{\bar{z},d}$ is the part of $\abar^0$ containing the
derivatives. As is shown in the appendix, the last two
equations are automatically satisfied in the parametrization
(\ref{par1}) and (\ref{par2}). Thus it remains to solve $S$
from (\ref{sig1}) and (\ref{sig2}). This can be done, resulting
in
\be \label{finals}
S=\deel{k}{2\pi}\int d^2 z \,\, \tr(
\prk(\adj{\abar^0})\prkd \Lambda^g) -\deel{k}{2\pi}
\int d^2 z \,\, \tr( \ll \dbar g g^{-1}) - S_{WZW}(g),
\ee
where $S_{WZW}$ is the Wess-Zumino-Witten action defined in
(\ref{wzw}). Equation (\ref{sig1}) follows straightforwardly
from (\ref{finals}), because varying $\prk\abar$ while keeping
$\pri A_z$ constant, means we only need to vary $\abar^0$, or,
equivalently, the $\mu_i$. Under such a variation,
\be
\delta S=\deel{k}{2\pi} \int d^2z \,\, \tr(
\delta (\prk\abar) \prkd \ll^g), \ee
showing (\ref{sig1}). To demonstrate (\ref{sig2}), we have to
vary $\pri A_z$ while keeping $\prk\abar$ constant. If we
denote the corresponding variation of $g$ by $\delta g$, we
find
\be \label{varvar}
\delta S =
\act{\prk(\adj{\abar^0})\delta (\prkd \ll^g)}
-\act{\ll\delta(\dbar g g^{-1})} -\delta S_{WZW}(g).
\ee
Using $\prk={\bf 1}-\prid$ the first
term in (\ref{varvar}) can be written as
\ba \label{varvarvar}
\act{(\adj{A^0_{\bar{z},f}}+\adj{\dbar})\delta (\ll^g) -
\prid(\adj{A^0_{\bar{z},f}}+\adj{\dbar})\delta (\pri A_z)}.
\ea
The second term of the last expression is already what we want,
so we would like the remainder of $\delta S$ to vanish.
Using straightforward algebra, this remainder can be written as
\ba
\delta S &=&
-\act{F(\ll^g,\adj{A^0_{\bar{z},f}}+\adj{\dbar})
(\adj{\delta})} \nonu
&=& -\act{F(\ll,A^0_{\bar{z},f})(\delta g g^{-1})}. \label{the1}
\ea
Recall that we constructed $\abar^0$ in such a way that $\pri
F(\ll+W,A^0_{\bar{z},f}+A^0_{\bar{z},d})=0$; restricting this
to the piece containing no derivatives, gives
$\pri F(\ll,A^0_{\bar{z},f})=0$. As $\delta g
g^{-1}\in\prid\bg$, it follows immediately that (\ref{the1})
vanishes, proving the validity of (\ref{sig2}).

The wave functions $e^S\Psi$ obtained here are the analogue of
(\ref{psi}) in a different polarization. In fact, they should
be seen as the Fourier transform of (\ref{psi}) with respect to
$\prkd \abar$. The first term of (\ref{finals}) can be more or
less understood as arising from this Fourier transform.
Furthermore, we see that in (\ref{finals}) part of the WZW
action has survived in the form $S_{WZW}(g)$.
The whole action (\ref{finals}) bears an interesting similarity
with the wave functions introduced in \cite{witten1}.
The wave functions
$\Psi$ that solve the classical $W_N$-Ward identities can be
obtained from a constrained WZW model as in \cite{beroog,kj1}. Having
solved the zero-curvature equations, we can in the next section
proceed with the computation of inner-products in this
polarization.

\newsection{The Covariant Action}

As explained in section 2 we have
to compute the inner-product between two
wavefunctions $\Psi_1$ and $\bar{\Psi_2}$ to
obtain our first result for
the covariant action for $W_N$ gravity. The wavefunction $\bar{\Psi}$
can be constructed in a similar way as $\Psi$ was constructed
in the previous
section. Introducing gauge fields $B_z,B_{\bar{z}}$, where the
fields are now in $\prkd B_{z}$ and $\prid B_{\bar{z}}$, the
wavefunction $\bar{\Psi}$ takes the form:
\be \label{psibar}
\bar{\Psi}[\prkd B_z,\prid B_{\bar{z}}] =
e^{\bar{S}(\prkd B_z,\prid B_{\bar{z}})}
\bar{\Psi}[\bar{\mu}_2,\ldots ,\bar{\mu}_N].
\ee
Here $\prkd B_z$ and $\prid B_{\bar{z}}$ are parametrized by
\ba \label{bbbar}
\prkd B_z &=& \prkd (h B^0_z h^{-1} - \dif h h^{-1}), \nonu
\prid B_{\bar{z}} &=& \prid (h \bar{\Lambda} h^{-1} -\dbar h h^{-1})
\equiv \prid \bar{\Lambda}^h,
\ea
with $\bar{\Lambda}=\Lambda^t=\sum_{i=1}^{N-1}e_{i+1,i}$,
\be \label{projbzero}
\prkd B^0_z = \sum_{i=1}^{N-1}\bar{\mu}_{N+1-i}e_{N,i},
\ee
and the
other components of $B^0_z$ can be computed from the condition
$\prid F(B^0_z,\bar{\Lambda})=0$. In (\ref{bbbar}) $h \in \exp (\pri
sl(N,\re))$, and $\bar{S}$ appearing in (\ref{psibar}) is given by:
\be \label{barac}
\bar{S} = \deel{k}{2\pi} \int d^2z\,\, \tr(\prkd(h B^0_z h^{-1})
\prk \bar{\Lambda}^h) + \deel{k}{2\pi} \int d^2z \,\, \tr(\bar{\Lambda}
h^{-1}\dif h) - S_{WZW}(h).
\ee
The last ingredient we need in our construction of the covariant action
is the K\"{a}hler form. Since this K\"{a}hler form should establish
the Fourier transformation from $\pri A_z$ to $\prid B_{\bar{z}}$ and
from $\prkd B_z$ to $\prk A_{\bar{z}}$ (or vice versa), it
is given by:
\be \label{keler}
K(A,B) = \deel{k}{2\pi} \int d^2z \,\, \tr(\pri A_z \prid B_{\bar{z}}
- \prk A_{\bar{z}}\prkd B_z),
\ee
(where the minus sign follows from standard Fourier theory:\, if one
uses $e^{ipx}$ as integration kernel to transform from $x$ to $p$,
one should use $e^{-ipx}$ to go from $p$ to $x$).

Using the explicit form of the inner-product
\ba \label{inp}
\left< \Psi_1 \mid \Psi_2 \right> &=& \int D(\pri A_z)D(\prk
A_{\bar{z}})D(\prid B_{\bar{z}})D(\prkd B_z)\,
e^{S + \bar{S}+K}\bar{\Psi}_1[\bar{\mu}]\Psi_2[\mu] \nonu
&\equiv& \int D(\pri A_z)D(\prk A_{\bar{z}})D(\prid B_{\bar{z}})
D(\prkd B_z)\,e^{S_{cov}(A,B)},
\ea
we can now read off the covariant action for $W_N$ gravity.
Writing
\be \label{psimu}
\Psi [\mu_2,\ldots,\mu_N] = \exp -S_{W_N} (\mu),
\ee
this result reads:
\ba \label{wnac}
S_{cov} &=& \deel{k}{2\pi}\int d^2z\,\,\tr(\prk (g^{-1}A^0_{\bar{z}}g)
\prkd \Lambda^g)
-\deel{k}{2\pi}\int d^2z\,\,\tr(\Lambda\dbar g g^{-1})-S_{WZW}(g)\nonu
  &+& \deel{k}{2\pi}\int d^2z\,\, \tr (\prkd (hB_zh^{-1})
\prk \bar{\Lambda}^h)
+\deel{k}{2\pi}\int d^2z\,\,\tr(\bar{\Lambda}
h^{-1}\dif h)-S_{WZW}(h)\nonu
  &+& \deel{k}{2\pi}\int d^2z\,\,\tr(\pri A_z\prid B_{\bar{z}} - \prk
  A_{\bar{z}} \prkd B_z)
  - S_{W_N}(\mu) - \bar{S}_{W_N}(\bar{\mu}).
\ea

This action depends on $g,h,\mu_i,\bar{\mu}_i$ which amounts to
a total number of variables given by: $2N(N-1)+2(N-1)=2(N^2-1)$.
On the other hand, one expects $W_N$ gravity to be described by
the fields $g_{\mu\nu},d_{\mu\nu\rho},\ldots$, which are {\em
symmetric} tensors of rank $2,3,\ldots,N$, resulting in a total
number of degrees of freedom given by:
$3+4+\cdots+(N+1)=\deel{1}{2}(N-1)(N+4)$.
The discrepancy between the total number of degrees of freedom
in $S_{cov}$ as given in (\ref{wnac}), and the total number of degrees
of freedom in $W_N$ gravity can be partially resolved as
follows: (i) $S_{cov}$
is invariant under a $Sl(N-1,\re)\times \re$ symmetry group,
which can be used to
gauge away $(N-1)^2$ degrees of freedom,
(ii) in $S_{cov}$ there are {\em auxiliary} fields
(\ie\ fields which appear only algebraically in $S_{cov}$),
which can be eliminated by replacing them by their equations of
motion.

\newsubsection{Symmetries of the Action}

To investigate the symmetries of the above action (\ref{wnac})
we define another projection operator $\pry = \pri \prid$,
by decomposing an arbitrary element of $sl(N,\re)$ in the
following way:
\be
sl(N,\re) = \prk sl(N,\re) \oplus \pry sl(N,\re) \oplus
\prkd sl(N,\re),
\ee
so an element $f \in \exp (\pry sl(N,\re))$ is of the form
\be \label{f}
f= \left( \begin{array}{cccc} {*} & \cdots & * & 0 \\
\vdots & & \vdots & \vdots \\
{*} & \cdots & * & 0 \\ 0 & \cdots & 0 & * \end{array}
\right) ,
\ee
and they form a $Sl(N-1,\re)\times \re$ subgroup.
We expect that $S_{cov}$ will be invariant under
transformations of the form
$g \rightarrow gf$ and $h \rightarrow f^{-1}h$.
(This is motivated by the fact that for the standard
polarization (where $\prk=0$ so $\pry={\bf 1}$) $f$ would be an
arbitrary element of $Sl(N,\re)$, and, as we saw in section 2,
for that case these transformations indeed leave the covariant action
invariant.)
So let us study how $S$ changes under the transformation
$g \rightarrow gf$ and $h \rightarrow f^{-1}h$, with $f$
as in (\ref{f}).

Note that for any element $f \in \exp (\pry sl(N,\re))$ we have
\be \label{ad}
Ad_f (\Pi_{i,k}sl(N,\re)) \subset \Pi_{i,k} sl(N,\re),
\ee
(and the same holds for $\Pi_{i,k}^{\dagger}$), where $Ad_f(g) =
f^{-1}gf$. One easily checks that this implies, that for any $X
\in sl(N,\re)$ we have $\Pi_{i,k} Ad_f (X) = \Pi_{i,k} Ad_f
(\Pi_{i,k}X) = Ad_f(\Pi_{i,k}X)$, \ie\
\be \label{comm}
\Pi_{i,k} \circ Ad_f = Ad_f \circ \Pi_{i,k}
\ee
(and again the same holds for the $\Pi_{i,k}^{\dagger}$).

Using (\ref{comm}) and the fact that $\prkd f^{-1}\dif f = 0$
we see that the first terms in the first and second line of
(\ref{wnac}) are invariant, whereas the rest of the action changes
as:
\ba \label{change}
\del S_{cov} &=& -\deel{k}{2\pi} \int d^2z\,\, \tr(g^{-1}\Lambda
g \dbar f f^{-1})-S_{WZW}(f)-\deel{k}{2\pi}\int d^2z\,\,
\tr(g^{-1}\dif g \dbar f f^{-1}) \nonu
& & -\deel{k}{2\pi}\int d^2z\,\,\tr(h\bar{\Lambda} h^{-1}\dif f f^{-1})
-S_{WZW}(f^{-1})+\deel{k}{2\pi}\int d^2z\,\, \tr(\dif f f^{-1}
\dbar h h^{-1}) \nonu
& & +\deel{k}{2\pi}\int d^2z\,\,\tr( \dif f f^{-1}\bar{\Lambda}^h
+\dbar f f^{-1}\Lambda^g + f^{-1}\dif f f^{-1}\dbar f),
\ea
where we used the Polyakov-Wiegman formula (\ref{powi}) and the
fact that $\tr(X\Pi_{i,k}Y)=\tr(Y\Pi_{i,k}^{\dagger}X)$.
Since
\be \label {resisnul}
S_{WZW}(f)+S_{WZW}(f^{-1})=\deel{k}{2\pi}\int d^2z \,\,
\tr( f^{-1}\dif f f^{-1}\dbar f),
\ee
it follows that $S_{cov}$ is invariant under the above
transformations.

In a similar way as was done in section 2 for the standard
polarization it is now straightforward to write down the action
in terms of the invariant product $G=gh$. We find
$S_{cov}=\Delta S-S_{W_N}(\mu)-\bar{S}_{W_N}(\bar{\mu})$, with
\ba \label{cvac}
\Delta S &=& \deel{k}{2\pi} \int d^2z \,\, \tr(\Lambda G
\bar{\Lambda} G^{-1}) + \deel{k}{2\pi} \int d^2z \,\, \tr(
\bar{\Lambda} G^{-1}\dif G) - \deel{k}{2\pi} \int d^2z \,\,
\tr (\Lambda \dbar G G^{-1})\nonu
&-& \deel{k}{2\pi} \int d^2z \,\, \tr(
\prk (A^0_{\bar{z}}-\bar{\Lambda}^G)\pry G
\prkd (B^0_z-\Lambda^G) \pry G^{-1})-S_{WZW}(G),
\ea
where we used the following decomposition for $G$:
$G=\prkd G \pry G \prk G$, which in terms of matrices looks like:
\be \label{Gontb}
G= \left( \begin{array}{cccc} 1 &  &  &  \\  & \ddots &  &  \\
 & & 1 & \\ {*} & \cdots & * & 1 \\  \end{array} \right)
\left( \begin{array}{cccc} * & \cdots & * & 0 \\ \vdots &  & \vdots
& \vdots \\ {*} & \cdots & * & 0 \\ 0 & \cdots & 0 & * \end{array} \right)
\left( \begin{array}{cccc} 1 & \cdots &  & * \\  & \ddots &  & \vdots \\
 &  & 1 & * \\  &  &  & 1 \end{array} \right) .
\ee
$S_{cov}=\Delta S-S_{W_N}(\mu)-\bar{S}_{W_N}(\bar{\mu})$
is our final result for the covariant action
for $W_N$ gravity. Note that in $S_{cov}$ only the $\prk$
part of $A^0_{\bar{z}}$ and the $\prkd$ part of $B^0_z$
appear. Thus specifying $A^0_{\bar{z}},B^0_z$ as in (\ref{projabarzero}),
(\ref{projbzero}) is sufficient to compute the action.
As we will see in the next subsection it
turns out that $S_{cov}$ still contains redundant degrees
of freedom.

\newsubsection{Auxiliary Fields}

At this stage our action depends on $G,\mu_i,\bar{\mu}_i$,
so we have reduced the number of degrees of
freedom to: $N^2-1+2(N-1)=(N+3)(N-1)$, which is still more
than one would naively expect for $W_N$ gravity.
Below we will further reduce this number due to the observation
that some fields only appear algebraically in the action, \ie\
auxiliary fields, which can in principle be eliminated by replacing
them by their equations of motion.
Unfortunately, the general expression (\ref{cvac}) valid for all
$Sl(N,\re)$ is not suitable for the determination of auxiliary fields.
Instead, we have to use an explicit parametrization for $G$
in (\ref{cvac}) in order to isolate the auxiliary fields.
We will work this out for the case of $Sl(2,\re)$ and $Sl(3,\re)$.
\\ \\
\underline{$Sl(2,\re)$}:
\\ \\
For this case we parametrize our $G$ as follows:
\be \label{gauss1}
G = \left( \begin{array}{cc} 1 & 0 \\ \omega & 1 \end{array}
\right) \left( \begin{array}{cc} e^{\phi} & 0 \\ 0 & e^{-\phi}
\end{array} \right) \left( \begin{array}{cc} 1 & -\bar{\omega} \\ 0 & 1
\end{array} \right).
\ee
Labeling $A^0_{\bar{z}},B^0_z$ as in (\ref{projabarzero}), (\ref{projbzero}),
respectively,
we find $S_{cov}=\Delta S - S_{W_2}[\mu]- \bar{S}_{W_2}[\bar{\mu}]$
with,
\newpage
\ba
\Delta S = -\deel{k}{2\pi}\int d^2z\Bigl[
\dif \phi \dbar \phi + \omega(2\dbar \phi +\dif \mu)
+\bar{\omega} ( 2\dif \phi + \dbar \bar{\mu}) \nonu
+\mu \omega^2 + \bar{\mu}\bar{\omega}^2 +
2\omega\bar{\omega}-(1-\mu\bar{\mu})e^{-2\phi} \Bigr],
\label{cosl2}
\ea
and $S_{W_2}[\mu]$ is the solution of the Ward-identity
\be
(\dbar - \mu\dif -2\dif \mu)\frac{\del S_{W_2}[\mu]}{\del \mu}
= -\deel{k}{4\pi}\dif^3 \mu. \label{cosl7}
\ee
{}From (\ref{cosl2}) we recognize that $\omega,\bar{\omega}$
are auxiliary fields. Replacing these fields by their
equations of motion gives:
\be \label{schematic}
S_{cov} = S_L[\phi,\mu,\bar{\mu}]+K[\mu,\bar{\mu}]-S_{W_2}[\mu]-
\bar{S}_{W_2}[\bar{\mu}].
\ee
Here
\be \label{lio}
S_L = \deel{k}{4\pi}\int d^2z \,\,
\sqrt{-\hat{g}}\left( \hat{g}^{ab}\dif_a \phi \dif_b \phi + 4e^{-2\phi}
+\phi\hat{R} \right),
\ee
is the well-known Liouville action, the metric $\hat{g}$ is defined
by $ds^2=|dz + \mu d\bar{z}|^2$, and $K[\mu,\bar{\mu}]$
reads
\be \label{bkano}
K[\mu,\bar{\mu}] = \deel{k}{4\pi}\int d^2z\,\,
(1-\mu\bar{\mu})^{-1}\left( \dif \mu \dbar \mu -\deel{1}{2}
\mu (\dbar \bar{\mu})^2-\deel{1}{2}\bar{\mu}(\dif \mu)^2\right).
\ee

So, finally, we have reduced our set of fields to the three
basic ones, namely the three components by which we label
the metric $g=e^{-2\phi}\hat{g}$. $S_{cov}$ as given in
(\ref{schematic}) is our final result for the case of $Sl(2,\re)$,
and is in fact almost equivalent to Polyakov's action for induced
2D gravity:
\be
S_{cov}=\frac{c}{96\pi} \int \!\!\! \int R\, \Box^{-1}R,
\ee
written out in components for the metric $g=e^{-2\phi}\hat{g}$.
The only difference is the cosmological term $\int d^2 z \,
\sqrt{-\hat{g}}e^{-2\phi}$, which one usually adds to the
induced action. The absolute value of the
coefficient in front of the cosmological
term is not important, as it can be arbitrarily rescaled by
adding a constant to $\phi$.

Note that the term in the Liouville action
(\ref{lio}) that is linear in $\phi$, is proportional
to the curvature $R$. This is due to the fact that the trace or
Weyl anomaly is also proportional to the curvature. Actually,
this fact can already be seen from (\ref{cosl2}), where we did
not yet eliminate $\omega$ and $\bar{\omega}$. The term linear
in $\phi$ in (\ref{cosl2}) is proportional to $\int \phi d\!\! \mbox{
\boldmath $\omega$}$, where $\mbox{\boldmath $\omega$}$ is the one form
$\omega\,dz-\bar{\omega}\,d\bar{z}$. This shows that $\mbox{\boldmath
$\omega$}$ should be interpreted as being the spin connection
\cite{herman}, since $R$ is the curvature of the spin connection.
\\ \\
\underline{$Sl(3,\re)$}
\\ \\
This case was already extensively discussed in \cite{jj}. There
it was shown that if we take the following Gauss decomposition for
$G$:
\be
G=\mat{1}{0}{0}{\omega_1}{1}{0}{\omega_3}{\omega_2}{1}
\mat{e^{\vp_1}}{0}{0}{0}{e^{\vp_2-\vp_1}}{0}{0}{0}{e^{-\vp_2}}
\mat{1}{-\bar{\omega}_1}{-\bar{\omega}_3}{0}{1}{-\bar{\omega}_2}{0}{0}{1},
\ee
and parametrize $A^0_{\bar{z}}$ and $B^0_z$ again as in
(\ref{projabarzero}) and (\ref{projbzero})
(with $\mu \equiv \mu_2,\,\nu\equiv \mu_3$),
the action contains $\omega_3,\bar{\omega}_3$ as auxiliary fields.
Substituting their equations of motion
results in the following action: $S_{cov}=\Delta
S-S_{W_3}[\mu,\nu]-\bar{S}_{W_3}[\bar{\mu},\bar{\nu}]$, with
\ba \label{finalres}
\Delta S = & &\hspace{-5mm}\deel{k}{2\pi}\int d^2 z\,\,\Bigl\{
       \hf A^{ij}\dif\vp_i\dbar\vp_j
      +\sum_i e^{-A^{ij}\vp_j}-A^{ij}(\omega_i+\dif \vp_i)
      (\bar{\omega}_j+\dbar \vp_j)\hspace{3cm}\\[.5mm]
  &   & \hspace{8mm}-e^{\vp_1-2\vp_2}(\mu-\hf \dif \nu -\nu\omega_1)
      (\mub+\hf \dbar \nub +\nub\bar{\omega}_1)-e^{-\vp_1-\vp_2}\nu\nub\nonu
  &   & \hspace{8mm}
      -e^{\vp_2-2\vp_1}(\mu+\hf \dif\nu+\nu\omega_2)(\mub-\hf \dbar\nub
      -\nub\bar{\omega}_2) +\mu T+\nu W + \mub \bar{T}
      + \nub \bar {W}\Bigr\}, \nonumber
\ea
where $A^{ij}$ is the Cartan matrix of $Sl(3,\re)$ $A^{ij}=
\left( \begin{array}{rr} 2 & -1 \\ -1 & 2 \end{array} \right)$.
$T,W,\bar{T},\bar{W}$ are defined
through the following Fateev-Lyukanov \cite{fatly} construction:
\ba
(\dif-\omega_2)(\dif-\omega_1+\omega_2)(\dif+\omega_1) & = &
\dif^3+T\dif-W+\hf \dif T, \nonumber \\
(\dbar-\bar{\omega}_2)(\dbar-\bar{\omega}_1+\bar{\omega}_2)
(\dbar+\bar{\omega}_1) & = & \dbar^3
+\bar{T}\dbar+\bar{W}+\hf \dbar \bar{T}, \label{fat}
\ea
and we shifted $\mu\rightarrow \mu-\deel{1}{2}\dif \nu,\,
\bar{\mu}\rightarrow \bar{\mu}+\deel{1}{2}\dif \bar{\nu}$.
The first part of $\Delta S$ is precisely a chiral $Sl(3,\re)$ Toda action,
confirming the suspected relation between $W_3$-gravity and Toda
theory, see also section 5.2.
Actually, one would expect that in a "conformal gauge",
the covariant $W_3$-action will reduce to a
Toda action. Indeed, if we put $\nu=\nub=0$ in $\Delta S$,
then also $\omega_1,\omega_2,\bar{\omega}_1,\bar{\omega}_2$
become auxiliary fields.
Substituting their equations of motion as well, we find that
\be
\Delta S=\deel{k}{4\pi}\int d^2 z \sqrt{-\hat{g}}
\left(\hf \hat{g}^{ab}\dif_a\vp_i\dif_b\vp_j A^{ij}+
4\sum_i e^{-A^{ij}\vp_j}+ R \vec{\xi}\cdot \vec{\vp}\right)
+4K[\mu,\mub], \label{covtoda}
\ee
where $K[\mu,\mub]$ is the same expression as for the $Sl(2,\re)$
case (\ref{bkano}), and $\hat{g}$ is again given by
$ds^2=|dz+\mu d\bar{z}|^2$. In the case of $Sl(3,\re)$,
$\vec{\xi}\cdot\vec{\vp}$,
with $\vec{\xi}$ being one half times the sum of the positive roots, is
just given by $\vp_1+\vp_2$.
The action (\ref{covtoda}) is the same Toda action that was originally
present in $\Delta S$ in a chiral form,
and the integration over $\omega_1,\bar{\omega}_1,\omega_2,\bar{\omega}_2$
has the effect of coupling it to a background metric $\hat{g}$.

Of course, the most interesting part of the action is
the part containing $\nu,\nub$. Unfortunately, if we
do not put $\nu=\nub=0$, we can integrate over either
$\ro_1,\rb_1$ or over $\ro_2,\rb_2$, but not over
both at the same time,
due to the presence of third order terms in $\Delta S$.
Another clue regarding the contents
of the action (\ref{finalres}) can be obtained by treating the
second and third line in (\ref{finalres}) as perturbations
of the first line of (\ref{finalres}).
This means that we try to make an expansion in terms of
$\mu,\mub,\nu,\nub$.
The saddlepoint of the $\omega$-terms is at $\omega_i=-\dif\vp_i$
and $\bar{\omega}_i=-\dbar\vp_i$.
{}From (\ref{fat}) we can now see that $T,W,
\bar{T}, \bar{W}$ are, when evaluated in this saddle point,
the (anti)holomorphic
energy momentum tensor and $W_3$-field that are
present in a chiral Toda theory
\ba
T &=& -\hf A^{ij}\dif \vp_i \dif \vp_j - \vec{\xi} \cdot \dif^2
\vec{\vp}, \nonu
W &=& -\dif \vp_1 ((\dif \vp_2)^2+\hf \dif^2\vp_2-\dif^2\vp_1)
+\hf\dif^3 \vp_1 - (1\leftrightarrow 2), \label{todafields}
\ea
and similar expressions for $\bar{T},\bar{W}$.

This suggests that the full
action $\Delta S$ contains the generating functional for
the correlators of the energy-momentum tensor and the
$W_3$-field of a Toda theory, "covariantly"
coupled to $W_3$-gravity. The presence of the third order terms in
$W,\bar{W}$ in (\ref{finalres}) prevents us from computing the
action of this covariantly coupled Toda theory. The same
structure is also present in the action for $W_N$ gravity, as we
will discuss in section 5.2.

In a similar way as for $SL(2,\re)$, the terms linear in $\vp_i$
in (\ref{finalres}) are expected to be related to the Weyl and
$W_3$-Weyl anomaly. The term linear in $\vp_i$ in
(\ref{covtoda}) shows that the Weyl anomaly is related to
shifts in $\vec{\xi}\cdot\vec{\vp}=\vp_1+\vp_2$. This suggests that the
$W_3$-Weyl anomaly is related to shifts in directions orthogonal
to $\vec{\xi}$, that is, to shifts of $\vp_1-\vp_2$. Writing the
terms in (\ref{finalres}) linear in $\vp_i$ as
\be \deel{k}{4\pi}\int \,(\vp_1+\vp_2)d\omega_+ +
3(\vp_1-\vp_2)d\omega_-,\ee
where the one forms
$\omega_+$ and
$\omega_-$ are given by
\be \omega_{\pm}  =  (\omega_1 \pm \omega_2)dz
- (\bar{\omega}_1 \pm \bar{\omega}_2) d\bar{z},
\ee
we see that $\omega_+$ plays the role of the spin connection,
whereas $\omega_-$ is some kind of $W_3$-spin connection, whose
curvature is presumably related to the $W_3$-Weyl anomaly. These
statements may acquire a more precise meaning when comparing
(\ref{finalres}) with a one-loop computation for the induced action
of $W_3$-gravity, starting with the action in \cite{kj2}.

\newsection{Properties of the Covariant Action}

At this stage the reader may well wonder, what all that we have
done so far has to do with covariant $W_N$-gravity. This is
probably best explained by looking at the case of ordinary
two-dimensional quantum gravity, and before proceeding with
general $W_N$-gravity, we will first discuss this much simpler and
better understood case.

\newsubsection{2D Quantum Gravity}

Given any action $S(g_{ab},\vp)$, where $\vp$ denotes some set
of matter fields, that is both invariant under
Weyl transformations $g_{ab}\rightarrow e^{\rho}g_{ab}$ and
under general co-ordinate transformations, one can define an
induced action for 2D gravity
\be \label{induc} e^{-S_{ind}(g_{ab})}=\int D\vp
e^{-S(g_{ab},\vp)},
\ee
by integrating out the matter fields $\vp$. If the theory had
no anomalies, $S_{ind}$ would reduce to an action defined on
the moduli space of Riemann surfaces. However, it is well known
that there are anomalies, resulting in a non-trivial
$g$-dependence of $S_{ind}$. The precise form of these
anomalies depends, of course, on the choice of regularization
scheme used in performing the path integral over the matter
fields. A scheme often used in
conformal field theory is $\zeta$-function regularization
\cite{zetaguys}. This
is a diffeomorphism, but not Weyl invariant regularization
method.
In \cite{bekn} it was shown
that if $S$ is the action for a $b-c$ system
of spin $j$, and one parametrizes $g$ via $ds^2=e^{\rho}|dz+\mu d
\bar{z}|^2$, one can compute
$\vars{S_{ind}}{\rho}$, which is proportional to the Liouville
action, and $\frac{\delta^2 S_{ind}}{\delta \mu\delta\mub}$,
which is also non-vanishing, due to the lack of holomorphic
factorization of $S_{ind}$. We will call this the holomorphic
anomaly.

In Chern-Simons theory, the holomorphic wave-functions contain
$\Psi[\mu]$, and the anti-holomorphic wave-functions contain
$\bar{\Psi}[\mub]$, which are both solutions of the Virasoro
Ward-identities. These wave functions can only carry a
holomorphic or anti-holomorphic diffeomorphism anomaly, but not
the two anomalies of the type mentioned before. Therefore,
these wave-functions do not fit naturally in the
$\zeta$-function regularization scheme, but in another where
the only anomalies are the holomorphic and anti-holomorphic
diffeomorphism anomaly. However, it is known that these two
schemes are related to each other via a local counterterm
$\triangle \Gamma$ \cite{stora}. This counterterm consists of two
pieces, a Liouville action and another term $K$, to cancel the
holomorphic anomaly, which is proportional to (\ref{bkano}).
Thus
\be \label{ohyes}
e^{-S_{ind}-\triangle \Gamma}=e^{-S(\mu)-\bar{S}(\mub)}.
\ee
All of this strongly suggests that the action $\Delta S= S+\bar{S}+K$
occurring in the inner-product
(\ref{inp}) is, in the case of $SL(2,\re)$, nothing but
the local counterterm $\triangle \Gamma$. That this is indeed the
case was shown in section 4.1 (see \cite{herman}):
upon integrating over the
nonpropagating fields $\omega,\bar{\omega}$,
the covariant action becomes precisely
equal to the local counterterm $\triangle \Gamma$. To make the
connection more precise, if we write $\Psi(\mu)=e^{-S(\mu)}$,
then (\ref{ohyes}) can be rewritten as
\be \label{ohyes2}
e^{-S_{ind}}=e^{\triangle \Gamma}\Psi(\mu)\bar{\Psi}(\mub), \ee
and we see that the partition function of induced gravity is
just the inner product as computed in Chern-Simons theory.
This also explains the name `covariant', because clearly we
have not fixed any gauge in (\ref{ohyes2}). For $W_N$, we
expect that a similar picture exists, although we have no
concrete realization of it. What we do have is the covariant
action, and before discussing possible implications this action
has for $W_N$-gravity, we will first study this action in some
more detail.

\newsubsection{Relation to Toda Theory}

In the case of ordinary gravity, the covariant action
contains the Liouville action. The natural generalization of the
Liouville action is the Toda action based on $sl(N,\re)$, which
is known to be deeply related to $W_N$-algebras \cite{toda}.
Indeed, we will now show that
the covariant action is closely related to Toda theory. The
cases $N=2,3$ were already dealt with in section 4, and we will
now consider the general case.
The relation to Toda theory is most easily established by putting
$\mu_i=\mub_i=0$. The action is then given
by (see (\ref{cvac}))
\ba S & = & \act{\ll G \bar{\ll} G^{-1}}+\act{\bar{\ll}
G^{-1}\dif G}-\act{\ll\dbar G G^{-1}} \nonumber \\
& & -S_{WZW}(G)-\act{\prk\bar{\ll}^h\prkd\ll^g}, \label{redac}
\ea
where $G=gh$, as explained in section 4. Under a variation of
$h$
\be \label{varh}
\delta S = -\act{(h^{-1}\delta h) F(\ll^G - h^{-1}\prkd(\ll^g)h,
\bar{\ll})}, \ee
and under a variation of $g$
\be \label{varg}
\delta S = -\act{(\delta g g^{-1}) F(\ll,\bar{\ll}^G-g\prk(\bar{\ll}^h)
g^{-1})}. \ee
Instead of $G=gh$ we will now use a Gauss decomposition $G=n_1bn_2$
for $G$, where $n_1$ is lower triangular, $b$ is diagonal, and
$n_2$ is upper triangular, and restrict ourselves to variations
of $n_1$ and $n_2$, for which (\ref{varh}) and (\ref{varg}) are
still valid. In terms of this Gauss decomposition, they can be
written as
\be \label{varn2}
\delta S = -\act{(n_2^{-1}\delta n_2) F(n_2^{-1}(b^{-1}\dif b)n_2 +
n_2^{-1}b^{-1}\pri(\ll^{n_1})bn_2, \bar{\ll})}, \ee
and
\be \label{varn1}
\delta S = -\act{(\delta n_1 n_1^{-1}) F(\ll,-
n_1(\dbar b b^{-1})n_1^{-1}-
n_1 b\prid(\bar{\ll}^{n_2}) b^{-1} n_1^{-1})}. \ee
The equations of motion for $n_2$ and $n_1$ read therefore
\be \label{mot2}
\Pi_< F(n_2^{-1}(b^{-1}\dif b)n_2 +
n_2^{-1}b^{-1}\pri(\ll^{n_1})bn_2, \bar{\ll})=0, \ee and
\be \label{mot1}
\Pi_> F(\ll,-
n_1(\dbar b b^{-1})n_1^{-1}-
n_1 b\prid(\bar{\ll}^{n_2}) b^{-1} n_1^{-1})=0, \ee
where $\Pi_<$ and $\Pi_>$ denote the projections on the space of
lower and upper triangular matrices respectively. The first
equation (\ref{mot2}) can be solved as follows: $\Pi_<
F(X,\bar{\ll})=0$ is certainly true if $X$ is upper triangular, and
$X$ is upper triangular if and only if $n_2 X n_2^{-1}$ is.
Applying this to $F$ in (\ref{mot2}), we see that the equations
of motion for $n_2$ are solved if $\pri(\ll^{n_1b})$
is upper triangular.
This is the
case if there exists an element $A\in\prkd\bg$ such that
\be \label{miu1}
n_1^{-1}\dif n_1 +n_1^{-1}\ll n_1 = A-\dif b b^{-1} +\ll. \ee
In a similar way, (\ref{mot1}) is solved if there is an element
$B\in\prk\bg$ such that
\be  \label{miu2} -\dbar n_2 n_2^{-1} +n_2 \bar{\ll} n_2^{-1}
= B + b^{-1}\dbar b +\bar{\ll}. \ee
If we now replace $A$ by $-n_1^{-1}An_1$ and $B$ by $-n_2 B
n_2^{-1}$, we see that $n_1$ is a gauge transformation relating
the connections $\ll-\dif b b^{-1}$ and $\ll+A$, and $n_2$ is a
gauge transformation relating $\bar{\ll}+b^{-1}\dbar b$ and
$\bar{\ll}+B$. These transformations are well known: they are the
Miura transformations that have been used \cite{fatly} to
produce free field expressions for $W_N$-algebras.
Here, the matrices $A$ and $B$ will contain these free field
representations. Let us now substitute the equations of motion
for $n_1$ and $n_2$ back into the full covariant action
(\ref{cvac}). After some manipulations, it reads
\ba
S & = & \deel{k}{4\pi} \int d^2 z \,\, \tr (b^{-1}\dif b
b^{-1}\dbar b)+\act{b\bar{\ll}b^{-1}\ll} -\act{\abar^0 A}\hspace{1cm}
\nonu &-&\act{ B^0_z B}-\act{\prk(n_1^{-1} \abar^0
n_1) b \prkd ( n_2 B^0_z n_2^{-1} )b^{-1} }.
\label{todd} \ea
The first two terms are just an expression for a Toda
theory in a flat background metric; the more conventional Toda
action follows immediately by substituting $b=\exp \mbox{\rm
diag}(\phi_1,\phi_2-\phi_1,\ldots,\phi_{N-1}-\phi_{N-2},-\phi_{N-1})$.
The first two terms read, when expressed in terms of $\phi_i$,
\be
\deel{k}{4\pi} \int d^2 z \,\, A^{ij} \dif \phi_i \dbar \phi_j +
\deel{k}{2\pi} \int d^2 z \,\, \sum_i e^{-A^{ij}\phi_j},
\ee
where $A^{ij}$ denotes the Cartan matrix of $sl(N,\re)$.
The third and fourth term of (\ref{todd})
show that, to lowest order, $\mu_i$
and $\mub_i$ couple simply the fields $W_i$ and $\bar{W}_i$, as
one would construct them from the Toda theory. The last, mixing
term in (\ref{todd}) has no simple interpretation in the Toda
theory.

One can in principle go through the same exercise with $\mu_2$
and $\mub_2$ unequal to zero; this was worked out in \cite{jj}
for the case of $W_3$, see also sect. 4.2.,
and the result in that case is that the
Toda theory gets coupled to a non-trivial background metric,
determined by $\mu_2$ and $\mub_2$. We expect that the same
thing is true for general $W_N$ algebras, although we have not
tried to do the computation. If one also puts other $\mu_i$ or
$\mub_i$ unequal to zero, it is much more difficult to solve the
equations of motion for $n_1$ and $n_2$ in full generality, and
we have been unable to do so.

{}From the previous calculations, one might be tempted to conclude
that the covariant action is just the generating function for
the correlators of the $W_i$ and $\bar{W}_i$ fields in a Toda
theory. This is, however, not true, because most of the fields
in $n_1$ and $n_2$ are not just auxiliary fields, and one cannot
therefore in general just substitute their equation of motion
back into the action. We will make a few more comments about
this in of the next section.

\newsubsection{$W$,$\bar{W}$-Invariance of the Covariant Action}

There is an interesting analogy between the covariant action for
$W_N$-gravity, and the covariant action given in
(\ref{inprod2}). If we start with an action $S=\int d^2z \,
\bar{\psi} \gamma^{\alpha} (\dif_{\alpha}+A_{\alpha}) \psi$,
then we can repeat the story in section 5.1, and define and
induced action by integrating over the fermions \cite{pol2}. Now
$\triangle \Gamma=K(A_z,\abar)$ is a kind of `chiral anomaly', with $K$
given by (\ref{kahler}). However, from another point of view,
$K$ is needed to restore the gauge invariances (\ref{transf}). If
we adopt this point of view here, the covariant action would
arise as an action needed to restore $W$ and $\bar{W}$
invariance\footnote{The precise form of the covariant action
does, however, not seem to be completely fixed by this
requirement alone.}. Thus it seems natural to look for the action
of the $W$ and $\bar{W}$ algebra on the covariant action, and to
check whether the covariant action indeed restores $W$ and
$\bar{W}$ invariance.

It turns out that this invariance is indeed present, although
the expressions involved are rather cumbersome. We will,
therefore, only describe the $W$-transformations, and omit the
tedious proof that these leave the full covariant action invariant.
We also will not give the $\bar{W}$-transformations, but they can
be easily written down, once the $W$-transformations are given.

First consider the wave-functions $\Psi(\mu_2,\ldots,\mu_N)$.
As we discussed in section 3.1, the Ward identities that
annihilate these wave functions are $F(\ll+W,\abar^0)\Psi=0$.
When acting on $\Psi$, $W_i$ is given by
$\frac{2\pi}{k}\vars{\Psi}{\mu_i}$. If we substitute these
expressions for $W_i$ back into $\ll+W$ and $\abar^0$, we will
denote the resulting expressions by $\ll+\wips$ and $\abaps$.
The Ward identities are then simply
\be \label{ward4} F(\ll+\wips,\abaps)=0. \ee
On the other hand, the counterterm $S+\bar{S}+K=\triangle S$
induces also certain
fields $W_i$, obtained by differentiating it
with respect to $\mu_i$. The corresponding matrix $W$ containing
these $W_i$ will be denoted by $W_{ind}$ and is easily found
from (\ref{wnac})
\be \label{wind} W_{ind}=-\frac{2\pi}{k}
g\prkd(g^{-1}\dif g+\adj{\ll}
-hB^0_zh^{-1})g^{-1}. \ee
In the same way as we constructed $\abar^0$ in section 3.1, we
can construct an $X\in\bg$ such that
\be \label{condition} \pri F(\ll+W_{ind},X)=0, \ee
once we specify $\prk X$. If
\be \label{condition2} \prk X =
\left( \begin{array}{cccc} {0} & \cdots & 0 & \epsilon_N \\
{0} & \cdots & 0 & \epsilon_{N-1} \\ \vdots & & \vdots & \vdots \\
{0} & \cdots & 0 & \epsilon_2 \\ {0} & \cdots & 0 & 0 \end{array}
\right), \ee
we will denote the corresponding solution $X$ of
(\ref{condition}) by $X(\epsilon)$. We now define the
following transformation rules for $G$ and the $\mu_i$:
\ba \label{trafg} \delta_{\epsilon}G & = & X(\epsilon)G, \\
\delta_{\epsilon}(\prk \abaps) & = & -\prk(\dbar X(\epsilon) +
[\abaps,X(\epsilon)]), \label{trafmu}
\ea
where the $\epsilon_i$ are the parameters of the
$W_i$-transformations. One can prove that (i) these
transformations form a $W_N$ algebra, and (ii) that these
transformations leave the inner product invariant. The
transformation rules (\ref{trafg}) and (\ref{trafmu}) look like
ordinary gauge transformations, although in this case $X$ is
field dependent. All this is closely related to the well-known
fact that $W$-transformations can be realized as
field-dependent $sl(N,\re)$ gauge transformations
\cite{beroog,pol3}; more
precisely, they are the gauge transformations that preserve the
form (\ref{azero}) of a connection. Here, the same mechanism is
working, although in a different setting. For instance, one of
the curious features of the transformation rule (\ref{trafmu})
is that $\delta_{\epsilon}\mu_i$ can contain
$\vars{\Psi}{\mu_j}$. Only for $W_2$ this is not the case. In
that case
the relatively simple transformation rules, leaving invariant
the covariant $W_2$ action, see (\ref{cosl2}) and
(\ref{cosl7}),
are given by
\ba
\delta_{\epsilon}\mu & = & -\dbar{\epsilon} - \epsilon\dif\mu +
\mu \dif\epsilon, \nonu
\delta_{\epsilon}\phi & = & \hf\dif\epsilon +\epsilon \omega, \nonu
\delta_{\epsilon}\bar{\omega} & = & -\epsilon e^{-2\phi}, \nonu
\delta_{\epsilon}\omega & = & -\hf\dif^2\epsilon +\mub
\epsilon e^{-2\phi}-\epsilon
\dif\omega -\omega\dif\epsilon.
\ea
In $\delta_{\epsilon}\mu$ one recognizes the transformation
rule for a Beltrami differential under an infinitesimal
co-ordinate transformation.
In fact, after substituting the equations of motion for
$\omega,\bar{\omega}$, we have
\ba \label{voorspel}
\delta_{\epsilon}\mu & = & -\dbar{\epsilon} - \epsilon\dif\mu +
\mu \dif\epsilon, \nonu
\delta_{\epsilon}\phi & = & \hf \dif\epsilon
-\frac{\epsilon}{1-\mu\bar{\mu}}(\dif \phi+\hf\dbar\bar{\mu})+
\frac{\bar{\mu}\epsilon}{1-\mu\bar{\mu}}(\dbar \phi + \hf \dif \mu),
\ea
which we expect to be the ordinary transformation
rules for the components of the metric $g$, defined by
$ds^2=e^{-2\phi}|dz+\mu d\bar{z}|^2$, under
general co-ordinate transformations.
Under the transformation
$x^{\mu}\rightarrow x'^{\mu}=x^{\mu}-\xi^{\mu}$
the metric changes as:
\be \label{gezeur}
\del g_{ab}=-g_{bc}\dif_a \xi^c -g_{ac}\dif_b \xi^c - \xi^c\dif_c g_{ab}.
\ee
Writing $\xi^a=(\xi,\bar{\xi})$, this implies the following
transformation rules for the components of the metric defined by
$ds^2=e^{-2\phi}|dz+\mu d\bar{z}|^2$:
\ba \label{meergezeur}
\del_{\xi,\bar{\xi}} \phi &=& \hf \dif \xi +\hf \dbar \bar{\xi} +\hf  \mu \dif
\xi
+\hf \bar{\mu} \dbar \bar{\xi}-\xi \dif \phi - \bar{\xi}\dbar \phi,\nonu
\del_{\xi,\bar{\xi}} \mu  &=& -\dbar \xi + \mu \dif \xi -\mu \dbar \bar{\xi}
-\xi \dif \mu -\bar{\xi} \dbar \mu + \mu^2\dif \bar{\xi},\nonu
\del_{\xi,\bar{\xi}} \bar{\mu} &=& -\dif \bar{\xi} + \bar{\mu}\dbar \bar{\xi}
-\bar{\mu}\dif \xi -\xi \dif \bar{\mu}-\bar{\xi}\dbar \bar{\mu}
+\bar{\mu}^2\dbar \xi.
\ea
Redefining our parameters as follows: $\epsilon = \xi+\mu\bar{\xi},\,
\bar{\epsilon}=\bar{\xi}+\bar{\mu}\xi$, the above rules become
\ba \label{nogmeergezeur}
\del_{\epsilon,\bar{\epsilon}} \phi &=& \hf\dif \epsilon +
\hf\dbar \bar{\epsilon}-
\frac{\epsilon-\mu\bar{\epsilon}}{1-\mu\bar{\mu}}(\dif \phi + \hf\dbar
\bar{\mu})
-\frac{\bar{\epsilon}-\bar{\mu}\epsilon}{1-\mu\bar{\mu}}
(\dbar \phi +\hf \dif \mu) ,\nonu
\del_{\epsilon,\bar{\epsilon}} \mu  &=&
-\dbar \epsilon + \mu \dif \epsilon -\epsilon \dif \mu ,\nonu
\del_{\epsilon,\bar{\epsilon}} \bar{\mu} &=&
-\dif \bar{\epsilon} + \bar{\mu}\dbar \bar{\epsilon}
-\bar{\epsilon}\dbar \bar{\mu},
\ea
which are the same as in (\ref{voorspel}) for the chiral case
$\bar{\epsilon}=0$. To find the general case one should
of course look at the $\bar{W}$ analogue of (\ref{trafg})
and (\ref{trafmu}).

Having established some of the basic properties of the
covariant action, we will now discuss the implications the
covariant action has for $W$-gravity.

\newsection{Discussion}

One of the main gaps in our knowledge of $W$-gravity, is that
we do not know what the proper set of fields is, on which the
$W$-algebra acts. In other words, what is the counterpart of the
metric in the case of $W_N$-gravity? Naively, one might think
that one just has to add symmetric tensor fields of rank
$3,\ldots ,N$, as we already mentioned in section 4,
to produce fields with the right spin. One has,
however, not been able to perform this construction in full
detail, and it is still quite a mystery how such a construction
should work, in which presumably the conformal factors of the
tensor fields should play the role of Toda fields, quantization
gives rise to anomalies, the generalized Weyl anomaly gives
rise to the Toda action, etc. It is also conceivable that
certain auxiliary fields are needed to form a full
`$W$-multiplet', and that part of these auxiliary fields become
propagating on the quantum level. This shows that it is
difficult to count a priori the number of degrees of freedom of
$W$-gravity.

It is possible to give an upper limit for the number of degrees
of freedom of $W$-gravity, because the covariant action
certainly has all degrees of freedom in it. Although we worked
only up to lowest order in $1/c$, we expect that the higher
order corrections will essentially only give a field and
coupling constant renormalization, as is the case for $W_2$
\cite{poly} and seems to be the case for $W_3$ as well
\cite{vannie}. Therefore, we can just count the number of
degrees of freedom in the covariant action, and it is given by
$(N+3)(N-1)$. From this we should certainly subtract the number
of auxiliary fields in the covariant action. We do not know
what this number is for the general case, but for $W_2$ it is
two, and in the case of $W_3$ it is at least two (see \cite{jj}
and
section 4.2).
However, integrating out more than two fields results in this
case in a nonpolynomial action, whose precise meaning is rather
obscure. What is certainly {\it not} true, is that all the
off-diagonal elements of $G$ are auxiliary fields for $N>2$, so
that it is in general impossible to reduce the covariant action
to a Toda-like action. This happens only if we put the
off-diagonal elements of $G$ on-shell and $\mu_i=\mub_i=0$, as
explained in section 5.2.

If we compare the $W_3$-action of \cite{jj} with the action for
$W_3$ strings given in \cite{pope1}, we see that the two are
partially identical, except that the action in \cite{pope1} does
not have terms which couple $\mu_i$ and $\mub_i$, nor terms
involving exponentials, whereas our
action does have both such terms. This is due to the fact that their
action is describing a matter system with $W_3$ symmetry,
whereas our action describes the induced action for $W_3$-gravity.
This relation is similar to the relation between
the action for a free boson on the one
hand, and the Liouville action on the other hand.

Regarding the question `what is the moduli space related to
$W$-algebras', this approach strongly suggests it is just (a
component of) the space of flat $Sl(N,\re)$-bundles. We know
that in the standard polarization the partition function of
Chern-Simons theory can be written as the integral of some
density over the moduli space of flat $Sl(N,\re)$-bundles. The
partition function does, of course, not depend on the choice of
polarization chosen, and we can therefore in principle also
rewrite the partition function of $W_N$-gravity as the integral
of some density over the moduli-space of flat
$Sl(N,\re)$-bundles. This moduli-space should arise by looking
at the space of differentials $\{\mu_i\}$ modulo $W$-transformations,
but it is difficult to make this relation more precise.

Clearly, there are many problems left in this field, and we
hope we will come back to some of those in the near future.

\vspace{1cm}

{\bf Acknowledgements}

We would like to thank P. van Nieuwenhuizen, B. de Wit and E.
Bergshoeff for stimulating discussions and helpful comments.

This work was financially supported by the
Stichting voor Fundamenteel Onderzoek der Materie
(FOM).

\startappendix

\newsection{Appendix}

In this appendix we will give a derivation of (\ref{extr1}) and
(\ref{extr2}):
\ba
\frac{2\pi}{k}\var{\prk\abar} & = & \prkd(g^{-1}Wg) +
\mbox{{\rm derivatives not containing}}\,\,\, \var{\mu_i},
\label{extra1} \\
-\frac{2\pi}{k}\var{\pri A_z} & = &
\prid(g^{-1}A^0_{\bar{z},d}g) +
\mbox{{\rm derivatives not containing}}\,\,\, \var{\mu_i}.
\label{extra2}
\ea
First of all, we will derive an expression for $\abar^0$, which
was defined in section 3.1. An important object here is the
linear operator $L:\bg \rightarrow \bg$ defined as follows: one
easily verifies that $X\rightarrow \pri(\mbox{\rm ad}\ll(X))$
defines an invertible linear operator
$\prid\bg\rightarrow\pri\bg$; $L$ is the inverse of this map,
extended by 0 to an operator $\bg\rightarrow\bg$. We also
define an element $F$ of $\bg$ by
\be \label{deff} F=\sum_{i=2}^{N} \mu_i \ll^{i-1}. \ee
Then
\be \label{omeg1} A^0_{\bar{z},f}=\frac{1}{1+L\dif} F, \ee and
\be \label{omeg2} A^0_{\bar{z},d}=\frac{1}{1+L\dif} L[F,W], \ee
where $(1+L\dif)^{-1}$ means $\sum_{i\ge 0}(-L\dif)^i$. To see
this, substitute (\ref{omeg1}) and (\ref{omeg2}) into
(\ref{zerocur}):
\ba F & = & -\dbar W + \dif(1+L\dif)^{-1}F
+\dif(1+L\dif)^{-1}L[F,W] \nonumber \\ & &
+\lha \ll+W,(1+L\dif)^{-1}L[F,W]\rha . \label{zerocurf}
\ea
We have to show that $\pri F=0$. First consider the term
quadratic in $W$: $[W,(1+L\dif)^{-1}L[F,W]]$; by definition,
$\mbox{\rm Im}L\in\prid\bg$, hence
$(1+L\dif)^{-1}L[F,W]\in\prid\bg$. This implies
$[W,(1+L\dif)^{-1}L[F,W]]\in\prkd\bg$, (cf. (\ref{prop1})), and
this term does not contribute to $\pri F$. By similar reasoning
it follows that \be \pri[W,(1+L\dif)^{-1}F]=\pri[W,F]. \ee What
remains is
\be \pri F=\pri\left( (\dif+\mbox{\rm ad}\ll)(1+L\dif)^{-1}
(F+L[F,W])+[W,F]\right). \label{finter} \ee
The definition of $L$ shows that
\be \pri(\mbox{\rm ad}\ll(L(X)))=\pri X, \label{hulp0} \ee
and from this we derive that
\ba \pri\left( (\dif+\mbox{\rm ad}\ll)(1+L\dif)^{-1}X \right) &
= & \pri\left( \sum_{i\ge 0}(-1)^i L^i \dif^{i+1} X +\sum_{i\ge
0}(-1)^{i+1}L^i\dif^{i+1}X +\mbox{\rm ad}\ll(X) \right) \nonumber
\\ & = & \pri([\ll,X]). \label{hulp1} \ea
We can use this to evaluate (\ref{finter}):
\ba \pri F & = & \pri([\ll,F+L[F,W]]+[W,F]) \nonumber
\\ & = & \pri([\ll,F])=0, \ea
where the last line is a trivial consequence of the fact that
$F$ defined in (\ref{deff}) commutes with $\ll$. Finally,
observe that $\abar^0$ as defined here is of the required form
(\ref{abarzero2}). One can also try to compute an
expression for $\abar^0$ by explicitly computing all the entries
of the matrix $\abar^0$ \cite{bilal1}, but this is less suitable for
general computations as performed in this appendix, and
certainly more complicated.

Armed with the expressions (\ref{omeg1}) and (\ref{omeg2}), we
can compute the right hand sides of (\ref{extra1}) and
(\ref{extra2}). Our next task will be to compute the left hand
sides of (\ref{extra1}) and (\ref{extra2}). Recall that
\ba \label{para1} \pri A_z & = &  \pri(g^{-1}\dif g+g^{-1}\ll g),
\\ \label{para2} \prk\abar & = & \prk(g^{-1}\abar^0 g +
g^{-1}\dbar g ) = \prk(g^{-1}F g). \ea
The last line follows easily from the fact that for
$X\in\prid\bg$, $\prk(\adj{X})=0$, so that, in the expression
(\ref{para2}), we can always redefine $\abar^0$ with an
arbitrary $X\in\prid\bg$. From (\ref{para1}) and (\ref{para2}),
we find that the variations of $\pri A_z$ and $\prk \abar$ in
terms of $\delta g$ and $\delta F$ are given by
\ba \label{vara1} \delta(\pri A_z) & = &
\pri(\adj{\dif(\delta g g^{-1})}+\adj{[\ll,\delta g g^{-1}]}),
\\ \label{vara2} \delta(\prk \abar) & = &
\prk(\adj{\delta F} + \adj{[F,\delta g g^{-1}]}). \ea
Now, in general, given fields $f_{\alpha}$ in terms of other
fields $\phi_{\beta}$, we can express $\var{f_{\alpha}}$ in
terms of $\var{\phi_{\beta}}$, by taking the transpose of the
inverse of the matrix $(\delta f_{\alpha}  / \delta \phi_{\beta}
)$. Here, we must first invert (\ref{vara1}) and (\ref{vara2}),
and then take the transpose of the resulting expressions.
Starting with (\ref{vara1}), and using once more that for
$X\in\prkd\bg$, $\pri(\adj{X})=0$, (\ref{vara1}) can be written
as
\be \pri((\dif+\mbox{\rm ad}\ll)(\delta g g^{-1}))=\pri(g\delta(
\pri A_z)g^{-1}). \ee
{}From (\ref{hulp0}) and (\ref{hulp1}) we see that
$\pri((\dif+\mbox{\rm ad}\ll)(1+L\dif)^{-1}LX)=\pri X$. This
shows that
\be \label{pp1}
\delta g g^{-1} = (1+L\dif)^{-1}L(g\delta(\pri A_z) g^{-1}).
\ee
Proceeding in the same way with (\ref{vara2}), one finds
\be \label{pp2}
\delta (\prk F)=\prk(g\delta(\prk\abar)g^{-1}-\mbox{\rm ad} F (
(1+L\dif)^{-1}L(g\delta(\pri A_z)g^{-1}))).
\ee
For equations (\ref{extra1}) and (\ref{extra2}), we only need
the $\var{\mu_i}$, or, equivalently, the $\var{F}$  behavior of
$\var{\pri A_z}$ and $\var{\prk\abar}$, \ie\ we only need to
look at (\ref{pp2}). The transpose $A^T$
of an operator $A$ is in our
case defined by requiring that for arbitrary $\bg$ valued functions
$X$ and $Y$, the following identity holds:
\be \label{deftra}
\int d^2 z \,\, \tr (X(AY)) =
\int d^2 z \,\, \tr ((A^TX)Y). \ee
Among other things, this implies that $\dif$ and $L$ are
anti-symmetric, $\dif^T=-\dif$ and $L^T=-L$. The computation of
the transpose of the operators in the right hand side of
(\ref{pp2}) is now straightforward:
\ba &&
\int d^2 z  \,\, \tr\left(X
\prk(g\delta(\prk\abar)g^{-1}-\mbox{\rm ad} F (
(1+L\dif)^{-1}L(g\delta(\pri A_z)g^{-1}))) \right) = \nonumber
\\ &&
\int d^2 z \,\, \tr \left( \delta(\prk\abar)(\adj{\prkd X})
\right) \nonumber \\ &&
- \int d^2 z \,\, \tr \left( \delta(\pri A_z)
(g^{-1} ( L (1+L\dif)^{-1}[F,\prkd X] ) g ) \right). \label{sigm}
\ea
Using $\var{\prk \abar}=\frac{k}{2\pi} W$ we read off from
(\ref{sigm}) that, up to terms containing derivatives with
respect to $g$,
\ba \var{\prk\abar} & = & \frac{k}{2\pi} \prkd(\adj{W}),
\label{rho1} \\
\var{\pri A_z} & = & - \frac{k}{2\pi} \prid\left(g^{-1}
(L(1+L\dif)^{-1}[F,W])g\right). \label{rho2} \ea
The first equation is just (\ref{extra1}), and the second one is
(\ref{extra2}), as we see from (\ref{omeg2}). Therefore
(\ref{extra1}) and (\ref{extra2}) are automatically satisfied in
the parametrization (\ref{para1}) and (\ref{para2}), as claimed.

\end{document}